# Title: Automatic recognition of element classes and boundaries in the birdsong with variable sequences.

**Short title:** Automatic recognition of birdsong.

**Authors:** Takuya Koumura[1, 2] and Kazuo Okanoya[1, 3]*

**Affiliations:**

[1] Department of Life Sciences, Graduate School of Arts and Sciences, The University of Tokyo, Tokyo, Japan

[2] Research Fellow of Japan Society for the Promotion of Science

[3] Cognition and Behavior Joint Laboratory, RIKEN Brain Science Institute, Saitama, Japan

* Corresponding author

E-mail: cokanoya@mail.ecc.u-tokyo.ac.jp

# Abstract

Researches on sequential vocalization often require analysis of vocalizations in long continuous sounds. In such studies as developmental ones or studies across generations in which days or months of vocalizations must be analyzed, methods for automatic recognition would be strongly desired. Although methods for automatic speech recognition for application purposes have been intensively studied, blindly applying them for biological purposes may not be an optimal solution. This is because, unlike human speech recognition, analysis of sequential vocalizations often requires accurate extraction of timing information. In the present study we propose automated systems suitable for recognizing birdsong, one of the most intensively investigated sequential vocalizations, focusing on the three properties of the birdsong. First, a song is a sequence of vocal elements, called notes, which can be grouped into categories. Second, temporal structure of birdsong is precisely controlled, meaning that temporal information is important in song analysis. Finally, notes are produced according to certain probabilistic rules, which may facilitate the accurate song recognition. We divided the procedure of song recognition into three sub-steps: local classification, boundary detection, and global sequencing, each of which corresponds to each of the three properties of birdsong. We compared the performances of several different ways to arrange these three steps. As results, we demonstrated a hybrid model of a deep neural network and a hidden Markov model is effective in recognizing birdsong with variable note sequences. We propose suitable arrangements of methods according to whether accurate boundary detection is needed. Also we designed the new measure to jointly evaluate the accuracy of note classification and boundary detection. Our methods should be applicable, with small modification and tuning, to the songs in other species that hold the three properties of the sequential vocalization.

# Author summary

A lot of animal species communicate with sequential vocalizations. The clearest example is human speech, in which various meanings are conveyed from speakers to listeners. Other animals also show interesting behaviors using sequential vocalizations for attracting mates, protecting territories, and recognizing individuals. Studying such behaviors leads not only to understanding of spoken language but also to elucidation of mechanisms for precise control of muscle movements and perception of auditory information. In studying animal vocalization, it is not rare case that long duration of sound data spanning over days or months must be analyzed, leading to a need of automatic recognition of vocalizations. In analyzing sequential vocalization it is often necessary to accurately extract temporal information as well as its contents. Another thing that must be considered is the rule for sequencing vocal elements, according to which variable sequences of vocal elements are produced. In the present study we propose methods suitable for automatic recognition of birdsong, one of the most intensively studied sequential vocalizations. We demonstrated the effectiveness of machine learning in automatically recognizing birdsong with temporal accuracy. Also we designed the new method to evaluate temporal accuracy of the recognition results.

# Introduction

Sequential vocalizations, in which voices are produced sequentially, have been a target of wide variety of researches. This is not only because they include human spoken language, but also because they serve as excellent models for precise motor control, learning, and auditory perception.

Birdsong is one of the most complex and precisely controlled sequential vocalizations, and has been widely and intensively studied [1-3]. Birdsong, as well as most of other sequential vocalizations, has several distinct properties. First, usually a song is a sequence of discrete vocal elements (called notes) [4]. Thus, by grouping similar notes into a single class, it is possible to convert songs into symbol sequences of note classes. Notes in a single class are considered to be generated by the same set of commands in motor neurons, which leads to the similar patterns of muscle activation to the similar acoustic outputs [5-8]. It is also known that auditory stimuli of notes in the same class invoke similar activation patterns in the auditory brain areas [9]. Second, temporal structure of the song is precisely controlled. Neural activities that are time-locked to a particular timing in a particular note class have been found during song production and perception [10-13]. Other studies have shown that temporal patterns of birdsong are constructed upon multiple levels of variability from local to global ones [14,15]. Thus, in analyzing birdsong it is important to accurately extract timing information such as note onsets and offsets. Finally, notes are sequenced not randomly but according to a certain probabilistic rule. Usually rules for note sequencing are unique to individuals and acquired by learning [16-20]. This rule for note sequence production is called song syntax. Taken together, in analyzing birdsong it is important to group notes into classes, extract timing information, and consider song syntax.

In behavioral and physiological studies on animal vocalization it is not a rare case when vocalizations in several days (or months) are to be analyzed [21,22]. For example a Bengalese finch, one of the model species for sequential vocalization, typically sings for ten minutes to one hour totally in a day, consisting of 5-30 thousand notes (depending on individuals), resulting in tens of hours of songs including hundreds of thousand notes in several days to be analyzed. Therefore for efficient analysis of vocalizations an accurate and robust automatic recognizer is strongly desired.

As stated above, in many cases of studies on sequential vocalizations such as birdsong, it is important to extract temporal information as well as its contents. This is one big difference and difficulty in recognizing sequential vocalizations for biological research compared to ordinary human speech recognition for application purposes, in which usually the priority is to convert sound data into text sequences and thus word or phoneme boundaries are not very much important [23,24]. On the contrary, vocalizations in non-human animals usually consist of smaller number of element classes and their combination patterns compared to human spoken language, which makes recognition easier in terms of pattern complexity. Therefore developing automatic recognizers of sequential vocalizations specialized for biological purposes, not just blindly using the methods for human speech recognition, is important for further research on animal vocalization.

Several previous studies have performed automatic recognition of birdsong, using dynamic time warping (DTW) [25-27], a hidden Markov model (HMM) [26], or a support vector machine (SVM) [28]. In machine learning in general, it is crucial to construct good representations of the data that separate data classes well in the feature space. In the previous studies, sound spectrums [25-27], mel-frequency cepstral coefficients (MFCC) [26], a small set of acoustic features [29], and a large number of features including spectrums, cepstrums, and their changes [28] have been used to describe properties of songs. However, there is no good reason to use MFCC in birdsong recognition because it has been designed for human speech. Also, it is not known whether the specific features used in the previous studies are suitable for vocalizations of other species as well. Although a SVM can automatically select good features from a large set of features, the problem of considering the initial feature set still remains. The desirable methods are ones that can automatically extract good features from data without manually engineering a set of features.

In the present study, to fulfill the three requirements stated above, we employed a hybrid model of a multi-layered neural network (also called as a deep neural network, DNN) and an HMM, with which high performances have been achieved in human speech recognition [24]. A DNN is known to have a capacity to find good representations of data by machine learning [30], making it possible to achieve robust note classification. Specifically we used deep convolutional network to handle long continuous sounds [31-33]. An HMM is good at handling variable sequences produced according to probabilistic syntax rules. Note boundaries were detected either with an HMM or by thresholding of amplitude and duration of note and silent intervals.

Performances of the recognizers were evaluated by three-way cross-validation with two types of measures. First measure is the one derived from the Levenshtein distance, which evaluates the accuracy of note classification ignoring temporal information. This measure corresponds to the word error rate in human speech recognition. Second measure is a newly designed one, motivated by the requirement of precise recognition of element boundaries. This measure jointly captures the accuracy of both note classification and boundary detection by selecting the best matched intervals in the recognized sequences and the correct ones. We call these two measures the Levenshtein error and the matching error, respectively. In machine learning in general, the larger the training data set the more generalization ability is obtained. It has been also the case in the previous study on birdsong recognition [28]. In the current study we compared the recognition results trained on two and eight minutes of training data randomly selected from the non-validation data set (2/3 of the whole data). In one bird the recognizer was trained on four minutes of training data as well.

## Results

**Data sets**

In total songs in twelve birds were recorded. All notes in recorded songs were manually located and classified. Songs in two birds which had more than 1% of manually unrecognizable notes were discarded. Songs in the remaining ten birds were used for the following evaluation. The average ± standard deviation of the total song duration in ten birds was 40.7 ± 18.4 minutes. The number of total notes and the number of note classes was 17930.9 ± 8941.8 and 8.1 ± 3.9, respectively. Songs in each bird were individually processed because songs were largely different among birds.

**Three steps in birdsong recognition**

In this study we divided the automatic recognition of birdsong into three sub-problems, each of which corresponds to one of the three properties of birdsong stated in the introduction section. First, notes must be correctly located in the continuous sound recordings by detecting note boundaries (note onsets and offsets). We call this step "boundary detection". Second, each note must be classified into a given number of classes (or the class for background noise). We call this step "local classification". The combination of boundary detection and local classification is equivalent to object spotting or semantic segmentation in two-dimensional object recognition [34,35]. Finally, outputs of the local classification are sequenced according to given song syntax. We call this step

"global sequencing". In the global sequencing step misclassifications of the local classifier were corrected by top-down information of the song syntax. The local classification and the global sequencing step can be seen as a bottom-up path and a top-down path in the song recognition, respectively.

To accomplish the song recognition, several ways of combining or arranging these steps are possible. In this study, we compared three different arrangements (Fig. 1). In the first arrangement, boundary detection, local classification, and global sequencing were sequentially conducted (Fig. 1a). In the second arrangement, local classification was conducted before simultaneously conducting boundary detection and global sequencing (Fig. 1b). In the last arrangement, local classification and global sequencing were simultaneously conducted before the simultaneous boundary detection and (another) global sequencing (Fig. 1c). The last arrangement included global sequencing twice in different ways. Hereafter we call these three arrangements "boundary detection -> local classification -> global sequencing (BD -> LC -> GS)", "local classification -> boundary detection & global sequencing (LC -> BD & GS)", and "local classification & global sequencing -> boundary detection & global sequencing (LC & GS -> BD & GS)".

The source code is available at https://github.com/takuya-koumura/birdsong-recognition.

**Evaluation**

Accuracy of the recognition was evaluated using the following two measures. To evaluate the accuracy of note classification, the Levenshtein distance between the output label sequence and the corresponding correct label sequence was computed. The Levenshtein distance is the minimum number of operations of insertion, deletion, and replacement that are needed to convert one sequence into another. The actual measure used for the evaluation was the total Levenshtein distance divided by the total number of notes in the correct sequences. In this paper this measure is called the Levenshtein error.

The Levenshtein error is designed to measure the difference of two symbol sequences, but does not have a capacity to capture the difference of note boundaries. The straightforward measure for jointly evaluating the accuracy of note classes and boundaries might be the total length of the time points in which the recognizer assigned different classes from those in the correct sequences. However, such a measure cannot distinguish the case in which two successive notes with identical classes were incorrectly recognized as one long note from the case in which such two notes were correctly recognized as two notes. Obviously the former result should have a higher error than the latter. Another measure might be the sum of the distances from the note onsets in the output sequences to the onsets of the nearest notes with the same classes in the correct sequences plus the distances between the offsets. This measure is likely to work well as long as all note classes in one sequence exist in another. In this study, we devised the new measure for jointly evaluating the note classes and boundaries without any specific constraints (Fig. 2). First, for each correct note interval, an output note with the same class and with the longest overlap with the correct one, if any, was matched. The overlapped sections in the matched intervals were regarded as the correctly recognized note intervals. Next, for each silent interval in the correct sequence, the correctly recognized silent intervals were defined as the sections in which no classes were assigned in recognition. Finally, the total length of the correctly answered note and silent intervals was divided by the total length of the input sequences, and subtracted from one. Hereafter we call this measure as the matching error. With this matching error, we can evaluate both note classes and boundaries in accordance with intuitive differences.

**Song recognition in the BD -> LC -> GS arrangement**

In the BD -> LC -> GS arrangement, first, note boundaries were located as the intervals with amplitude and duration larger than certain thresholds (Fig. 3). The thresholds were determined to minimize the matching error in the training data set ignoring the note classes. Then sound spectrograms in a sliding time window were locally classified into a given number of note classes with a DNN (Fig. 4a). The width of the input time window was 111 ms, roughly corresponding to the duration of a single note. The architecture of the DNN was as follows, from input to output: an input layer; three sets of convolutional layers, cascaded cross channel parametric (CCCP) pooling layers, and max-pooling layers; a fully-connected layer; a softmax layer. See the materials and methods section for detailed configurations. The outputs of local classification were averaged in each detected note interval (Fig. 1a). Finally, averaged classification outputs were combined with the syntax with an HMM. The song syntax was modeled with a second-order Markov model in accordance with the previous study [36], in which the probability of the next note classes depends on the previous note and the second-previous note (Fig. 5a).

The average Levenshtein errors were as low as 1.58% with 2 minutes of training data and 1.33% with 8 minutes of training data (Table 1), ranging from 0.19% to 5.01% (Fig. 6a), whereas the average matching errors were 4.41% and 4.28% with 2 minutes and 8 minutes of training data, among which errors in one bird reached nearly 10%. These results suggest that note boundaries are not accurately detected by the BD -> LC -> GS arrangement. Indeed the recognition result in the bird with the highest matching errors shows that boundaries between notes with almost no silence could not be detected by the thresholding method (Fig. 7b). Obviously the matching errors are lower-bounded by the errors of boundary detection, measured with the matching errors ignoring note classes (Fig. 6c). In contrast, if boundary detection worked well, the following classification and sequencing were successful (Fig. 7a). Both the Levenshtein errors and the matching errors decreased as the length of the training data increased except for the Levenshtein errors in two birds.

**Song recognition in the LC -> BD & GS arrangement**

In the LC -> BD & GS arrangement, first, local classification was performed with a DNN which had the same architecture as that in the BD -> LC -> GS arrangement (Fig. 4a). Then the boundary detection and global sequencing steps were simultaneously conducted with an HMM. To accurately detect note boundaries with an HMM, each note was divided into three parts with the same duration (Fig. 8), and each hidden state in the HMM was divided into four, first three of which emitted three parts of a note and last of which emitted the silent interval between notes (Fig. 5b). The last two sub-states had the connections to the next note, which represented the transition from one note to another. Dividing each note into more than two parts was crucial. This is because if no division was made in a note, the HMM would not be able to distinguish two notes with a very small silent interval from one long note by mistakenly inserting or skipping short silences (notes A and C in Fig. 5b), and because if notes were divided into two parts, small misclassifications of two sub-divisions would cause unwanted onsets and offsets at the positions of the misclassifications. (notes B in Fig. 5c).

Examples of the recognition results show that sub-divisions in notes were correctly classified by the DNN, resulting in correct sequencing and boundary detection with the HMM (Fig.

9), even in the data poorly recognized with the BD -> LC -> GS arrangement (Fig. 9b). The average Levenshtein errors were 1.45% with 2 minutes of training data and 1.06% with 8 minutes of training data, and the average matching errors were 2.14% and 1.97% (Table 1). Both errors were lower than those in the BD -> LC -> GS arrangement. Especially the highest matching errors in the LC -> BD & GS arrangement was lower than the lowest in the BD -> LC -> GS arrangement (Fig. 6e), suggesting that the LC -> BD & GS arrangement is more suitable for recognizing precise temporal information. Both types of errors decreased as the length of the training data increased except for the Levenshtein errors in two birds.

**Song recognition in the LC & GS -> BD & GS arrangement**

In the LC & GS-> BD & GS arrangement, global sequencing as well as local classification was performed with a DNN. To include the syntax information that spanned over more than one notes in the DNN, additional fully-connected layer was inserted in the DNN, resulting in the width of the input time window widened to 303 ms, roughly covering three successive notes (Fig. 4b). The boundary detection and (another) global sequencing steps were the same as the LC -> BD & GS arrangement.

Examples of the recognition results show that recognition was accurately performed in the LC & GS -> BD & GS arrangement as well as in the LC -> BD & GS arrangement (Fig. 10). Both the average Levenshtein errors and the average matching errors were lower than those in the LC -> BD & GS arrangement, although the differences were small (Table 1, Fig. 6). These results demonstrated the capacity of a DNN to capture the syntax information ranging over multiple successive notes. Both types of errors decreased as the length of the training data increased except for the matching errors in one bird.

**Summary of the validation errors**

To summarize the performances of the recognizers, we compared the average validation errors over birds in each arrangement and training length (Table 1). Both types of errors were the lowest in the LC & GS -> BD & GS arrangement and the highest in the BD -> LC -> GS arrangement. Especially the matching errors in the BD -> LS -> GS arrangement were far higher than those in other two arrangements. The larger the training data, the lower both types of errors were, as has been shown in most of other machine learning studies. In all arrangements, the matching errors were lower than the Levenshtein errors. Although direct comparison of these two types of errors are not very much meaningful, probably the differences reflect the errors in boundary detection because the matching error is the measure for both note classification and boundary detection whereas the Levenshtein error is the measure for note classification only.

# Discussion

In the current study we evaluated the three different arrangements to automatically recognize songs in Bengalese finches, and with two arrangements achieved sufficiently low validation errors (~ 2%) for practical use in biological studies (Tabel 1). All arrangements used a DNN for local classification and an HMM for global sequencing step, demonstrating the effectiveness of the hybrid use of a DNN and an HMM for recognizing birdsong as well as human speech in the previous studies [24,37]. To the best of our knowledge this is the first time that a hybrid DNN/HMM model was applied to automatic recognition of birdsong.

The matching errors were higher in the BD -> LC -> GS arrangement than in the other two arrangements, suggesting that boundary detection should be performed with an HMM rather than by amplitude and duration thresholding. The Levenshtein errors were also higher in the BD -> LC -> GS arrangement but were acceptably low for practical use (Table 1). One advantage of the BD -> LC -> GS arrangement is that by abandoning boundary detection with an HMM the number of target classes in the DNN decreases approximately by a factor of three, making the computation for classification faster. The faster the computation, the finer the parameters could be tuned, possibly leading to better generalization. Moreover, slightly better recognition in the LC & GS -> BD & GS step compared to that in LC -> BD & GS step revealed the power of a DNN in handling data with complex and hierarchical structure. Therefore we propose that either of the BD -> LC -> GS or the LC & GS -> BD & GS arrangement should be employed according to the objectives: when accurate note classification is the first priority and the information of note boundaries were not important, the BD -> LC -> GS arrangement should be used; when both accurate note classification and boundary detection are required, the LC & GS -> BD & GS should be used.

The recognition methods investigated in this study should be applicable in all kinds of studies on animal vocalization with variable sequences that requires accurate element classification and/or element boundary detection. Essentially these methods do not depend on the particular features in acoustic data because either a DNN and an HMM is not specialized to particular forms of inputs. Especially a DNN is known to be good at learning good features from data without manual feature engineering [30,32], and with an HMM syntax information of variable sequences can be incorporated into recognition process.

Another achievement in this study is designing of the matching error, by which recognition results of both note classes and note boundaries can be evaluated. This measure can also be used for evaluating note boundary detection without classification by setting all classes of notes to identical (ie. grouping all notes into a single class).

**Other techniques that could possibly improve the results**

There are several techniques on the DNN that could possibly decrease the validation errors. One of them is the drop-out technique, in which at each iteration of the training a certain portion of randomly chosen network nodes are turned off [38]. This procedure can be seen as training multiple networks at the same time and using average outputs of them in the recognition phase. In the current study outputs of networks trained on three different data sets in the training data were averaged in recognition. Previous studies on in image recognition have shown that averaging outputs of multiple networks with different architectures improves generalization, although it will take computational time proportional to the number of the networks [39]. For parameter updating of the DNN, using the resilient propagation method, in which the amount of parameter changes is computed adaptively, might result in more robust parameter learning [40,41]. There is also a technique to virtually increase the amount of training data by artificially distorting the data [33,39]. For image recognition some operations for data distortion such as adding Gaussian noise, scaling, shearing, and local distortion are known to increase generalization. Finding the effective data distortion for acoustic recognition would be a helpful future study. It is not known, for example, which of these operations are effective: local scaling along time axis, along frequency axis, or both.

In the current study we described the song syntax in Bengalese finches with a second-order Markov model in accordance with the previous study [36]. However, there are other models that have been proposed to be suitable in describing song syntax in Bengalese finches such as an HMM [42], a simplified renewal process [43], and a k-reversed automaton [44]. Song syntax in canaries is well described by a prediction suffix tree [19]. It might result in more accurate recognition if such syntax models are used in the global sequencing step. A k-reversible model and a prediction suffix tree can be easily implemented in our HMM framework. The simplified renewal process can be implemented as well if the number of repetition of a single note class is limited. To use a HMM as a syntax model in the global sequencing step that already uses an HMM for sequencing, a hierarchical HMM could be considered.

Joint training of the DNN and the HMM using discriminative forward training [33] or maximum mutual information [45] could also improve the recognition results, although the effectiveness of the joint training is not sure when the correct note intervals were given as in the case of this study.

In the current study the 0th order discrete prolate spheroidal sequences (DPSS) was used as the taper of the short time Fourier transform to compute the sound spectrograms. Using average spectrum of the multiple tapers could generate more robust spectrograms against background noise [29,46].

**Hyper-parameters that were not tuned**

There were a lot of hyper-parameters in the training procedures. Since it is virtually impossible to tune all of those parameters due to the constraint of time for training, the values of some hyper-parameters were presumably fixed in the current study. Thus, finely tuning such hyper-parameters may improve the results. In training the DNN, learning rate of the gradient decent method was decreased by half once the classification score stopped improving on a portion of the training data. The decrease ratio of half was arbitrary. Thus better training schedule could exist. Also the architecture of the DNN such as the number of layers, the filter size of the convolutional layers, or the size of the fully-connected layers was pre-fixed.

Other untuned hyper-parameters were the batch size for training the DNN, the size of the Fourier transform, the step width of the spectrograms, and the parameter for the 0th order DPSS. These hyper-parameters could be tuned to minimize the recognition error in cross-validation within training data sets. See the materials and methods section for the current settings of the hyper-parameters.

**Limitations and possible future directions**

One obvious limitation of the present study is that pre-determined note classes and boundaries are required to train the recognizer. Although the length of songs required for training data sets is very short (~ 2 minutes), the preparation of them might be troublesome if there are tens of birds to be analyzed. To solve this problem, currently we are trying to establish the methods for unsupervised training or note clustering with features extracted by generative DNNs such as deep generative stochastic networks [47].

Another limitation is that in the current study songs were located manually from the whole sound recordings. This is because large part of the recordings were background noise that is not our current interest and recognizing recordings including hours of background noise takes a lot of computational time. Our methods are expected to have a capacity to locate notes in the whole recordings including long background noise, but this capacity needs to be evaluated properly in the future work. Another possible way to locate songs may be the similar method as this study assigning a single class to the whole songs and another class to the between-song silent intervals. Perhaps strict temporal resolution is not required in this song locating phase, and thus data could be down-sampled to fasten the computation.

In this study we only recognized songs in Bengalese finches. With small modifications and tunings, our methods are expected to work well in sequential vocalizations in other species because most sequential vocalizations have the three properties introduced in this paper: element classifiability, importance of timing information, and probabilistic sequencing rules. Currently we are evaluating similar recognition methods in vocalizations in other species such as other songbird species, rodents, and gibbons. However, there might be more suitable methods for songs in rodents that consist of vocal elements with long frequency-modulated sound. Also similar methods may be applied to recognizing vocalization in human babies to automatically extract both contextual information and timing information [48].

## Materials and Methods

### Ethics Statement

The experimental procedure were approved by the Institutional Animal Care and Use Committee of the University of Tokyo.

### Data acquisition

An adult male Bengalese finch (day post hatch > 120 days) was put into a sound attenuation chamber. After a habituation period of at least two days to the recording environment, sound was recorded during 14 hours of light on interval using a microphone (PRO 35, Audio-Technica Corporation, Japan), an amplifier (MicTube Duo, Alesis, United States), and an audio interface (OCTA-CAPTURE, Roland, Japan) with 32 kHz sampling rate. Light on and off intervals (14h and 10h, respectively) were controlled by an LED light. Food and water were given ad libitum. Songs in twelve birds were recorded.

**Data sets**

All sequential vocalizations in the recorded sound were manually extracted by visual inspection of the sound spectrogram. Sound spectrograms were computed using short time Fourier transform with a size of 512 and a step of 32 (corresponding to 1 ms), in which frequency band between 1 and 8 kHz was used in all of the following computations. In computing spectrums, the 0th order DPSS with a parameter $W = 4 / 512$ was used as a taper [46]. Spectrograms were mean-subtracted and divided by the standard deviation. The means and the standard deviations were computed in the training data sets (defined below).

Notes were located and classified manually by visual inspection of the spectrograms. The objectivity of this manual note locating and classification was to some extent guaranteed by the low cross-validation errors shown in this paper. Non-singing calls were labeled into one class. Occasionally there were manually unclassifiable notes such as ones which do not appear to belong to any classes or which have intermediate appearance of more than one classes. Classes with notes less than 1% of the total number of notes in the songs were also labeled as unclassifiable as well.

Note sequences, separated by non-singing calls, with more than seven notes and less than 300 ms silence between notes were extracted as songs. Birds with unclassifiable notes more than 1% of the total number of notes in the songs were discarded, keeping ten birds out of twelve recorded. Then to exclude the unclassifiable notes from the data sets each song was segmented by the unclassifiable notes. Segmented songs with less than three notes were discarded. Segmented songs with more than 15 notes were further segmented so that all sequences contained less than 16 notes because uneven length of sequences would lead to inefficient computation in terms of memory management and parallelization.

Note sequences were divided into three groups for three-way cross-validation. In machine learning in general, the larger the training data set the more generalization ability is obtained. It has been also the case in the previous study on birdsong recognition [28]. In the current study we compared the recognition results trained on two and eight minutes of training data sets randomly selected from the sequences in non-validation set (2/3 of the whole data set). Training using four minutes of training data was also conducted in one bird. Since the total sequence length differed among birds (ranging from 13.3 and 63.9 minutes), training using whole non-validation set was not performed. When a need of tuning hyper-parameters arises, the training data set was further divided into three to perform cross-validation within the training data. The hyper-parameters were set to the values that minimized the validation error in the cross-validation within the training data. Songs in each bird were individually processed because songs were largely different among birds.

**Boundary detection by amplitude and duration thresholding**

In the BD -> LC -> GS arrangement, note onsets and offsets were detected by amplitude and duration thresholding (Fig. 3). First, sound intervals with amplitude larger than a certain threshold were extracted as non-background intervals (orange bars in Fig. 3). Amplitude was computed as the sum of the logarithmic amplitude spectrum between the frequency band of 1 and 8 kHz in each 1 ms time bin of the spectrograms. Then among the extracted non-background intervals, those with silent intervals shorter than a certain threshold between them were concatenated. Finally, intervals with duration shorter than a certain threshold were discarded. The remaining intervals were considered as the note intervals (blue bars in Fig. 3). The three thresholds were determined to minimize the matching error ignoring the note classes among the training data set. As a result, the optimal threshold for the silent intervals was zero in all conditions in all birds, meaning that no two intervals were concatenated.

**Local classification with a deep convolutional network**

In all three arrangements, spectrograms within a fixed-length time window were classified into given note classes with a deep convolutional network [33] (Fig. 4). A deep convolutional network serves as a feature extractor and classifier [30-32]. In the BD -> LC -> GS and LC -> BD & GS arrangements the architecture of the network was as follows, from input to output: an input layer; three sets of convolutional layers, cascaded cross channel parametric (CCCP) pooling layers, and max-pooling layers; a fully-connected layer (size 240); a softmax layer (Fig. 4a). The filter size of the first two convolutional layers and the last convolutional layer were $5 \times 5$ and $4 \times 4$, respectively. The number of channels in the convolutional layers and the CCCP pooling layers was 16 in each. All convolutional layers, the CCCP pooling layers, and the fully-connected layer had rectified linear activation units. The filter size and the stride size of the max-pooling layers were both $2 \times 2$. The height of the input layer was 112, corresponding to the frequency band between 1 kHz and 8 kHz. The CCCP pooling layers can be seen as small networks acting as activation functions in the convolutional layers [49]. They were implemented by convolutional layers with an 1×1 filter size. The fully-connected layer was implemented by a convolutional layer with 240 channels. Thus the width of the input time window determined the filter width of the fully-connected layer. In the BD -> LC -> GS and the LC -> BD & GS arrangements the width of the input time window was 96 (111 ms), resulting in the filter width of the fully-connected layer to be 9. This width was determined so as to roughly cover duration of a single note. The filter height of the fully-connected layer was 11 to cover the whole input height of 112 (from 1 to 8 kHz). In the LC & GS -> BD & GS arrangement, another fully-connected layer was inserted before the softmax layer, hoping to capture the syntax information that spanned over more than one note (Fig. 4b). The filter width of the second fully-connected layer was 25, corresponding the input window width of 288 (303 ms), roughly covering the duration of three successive notes. Thus the aim of inserting this layer was to integrate the local single-note information in the lower fully-connected layer into the global syntax information over three notes. In other words, it tries to implicitly combine the outputs of local classification and the global syntax in the form of a trigram syntax model. To ensure proper classification under the lower fully-connected layer, the network without the upper fully-connected layer was trained before inserting the upper fully-connected layer and training the whole network.

Updating of parameters (weights and biases) was performed by a simple stochastic gradient descent method with cross-entropy cost function. The learning rate was determined in each training data set as follows. First the initial search of the learning rate was conducted using 2/3 of the training data set for training and the other 1/3 for validation. Training was conducted with various learning rates from 0.001 to 0.04. The initial learning rate was set to the value that achieved the lowest validation error in one of the first 32 training iterations. Then the full training was conducted on the 2/3 of the training data as long as the validation error in the other 1/3 kept decreasing. When the validation error stop decreasing, the learning rate was decreased by half and the training was continued. This procedure was repeated three times. The full training was performed on the three different combinations of the training and validation data, yielding three parameter sets in each training data set. In the recognition phase, the outputs of those three networks were averaged. The mini-batch for the training was selected randomly so that the total length in each mini-batch did not exceed 32 s.

The network weights and biases were initialized according to [50]. The initial weights were sampled from a Gaussian distribution whose mean was zero and standard deviation was the square root of two divided by the number of incoming connections to the particular node. All biases were initialized to zero. The random seed for the initial weights were searched at the same time with the initial search of the learning rate.

In the case of BD -> LC -> GS arrangement, the size of the softmax layer was the number of note classes. In the other two arrangements, to perform the following boundary detection with an HMM, the size of the softmax layer was three times the number of note classes plus one corresponding to the background noise.

**Boundary detection and global sequencing with a hidden Markov model**

In the BD -> LC -> GS arrangement, the outputs of local classification were combined with the global syntax information with a hidden Markov model (HMM) [24,37] (Fig. 1a). Generally when an HMM are combined with a DNN, the outputs of the DNN are considered as the posterior probabilities of the hidden states [24,51]. In the current study, song syntax was described with a second-order Markov model [36] (Fig. 5a). To describe the syntax of note sequences with a second-order Markov model, the structure of an HMM was modified so that the transition between hidden states were mediated by generation of a symbol corresponding to the note class, and that the outputs of the DNN were considered as the posterior probabilities of the symbols.

*next state ~ P*(*next symbol | current state*),

*DNN output = P*(*next symbol | spectrogram*).

The transition probability between hidden states was computed from the training data set with smoothing by adding a constant value before dividing by the sum. The smoothing constant was determined by cross-validation within the training data set. The transition probabilities from initial state and the next state of the initial state were assumed to be uniformly distributed. The outputs of the DNN were averaged over the each sound interval detected by thresholding to obtain the posterior probability of the hidden states in each interval. The optimal state sequences for the computed posterior probabilities were estimated by Viterbi algorithm, which were then converted into the label sequences of the note classes.

In the CL -> BD & GS and the CL & GS -> BD & GS arrangements, boundary detection was performed simultaneously with global sequencing (Fig. 1b & c). To accurately detect note boundaries, each note was divided into three parts with the same duration (Fig. 8), and each state was divided into four, first three of which emitted three parts of a note and last of which emitted the silent interval between notes (Fig. 5b). These divided sub-states are connected in the left-to-right manner including self-transitions. The last two sub-states had the connections to the next state, which correspond to the transition from one note to another. The transition probabilities in these transitions followed the second-order note transition probabilities in the training data set with smoothing. The other transition probabilities between sub-states were assumed to be uniformly distributed.

In general cases of recognition in an HMM with a DNN, the outputs of the DNN are considered as the posterior probability of the hidden states given the acoustic data. The posterior probability is often converted into the emission probability of the acoustic data given the hidden states by Bayes' rule [24,51]. In this study we chose whether to conduct this conversion according to the cross-validation within the training data sets.

**Computation**

All computations and sound recording were implemented in the custom written java program. The source code is available at https://github.com/takuya-koumura/birdsong-recognition. Training of the DNN and recognition were conducted using cuDNN library on graphic processors (GTX 970 or 980, NVIDIA, United States).

# Figures

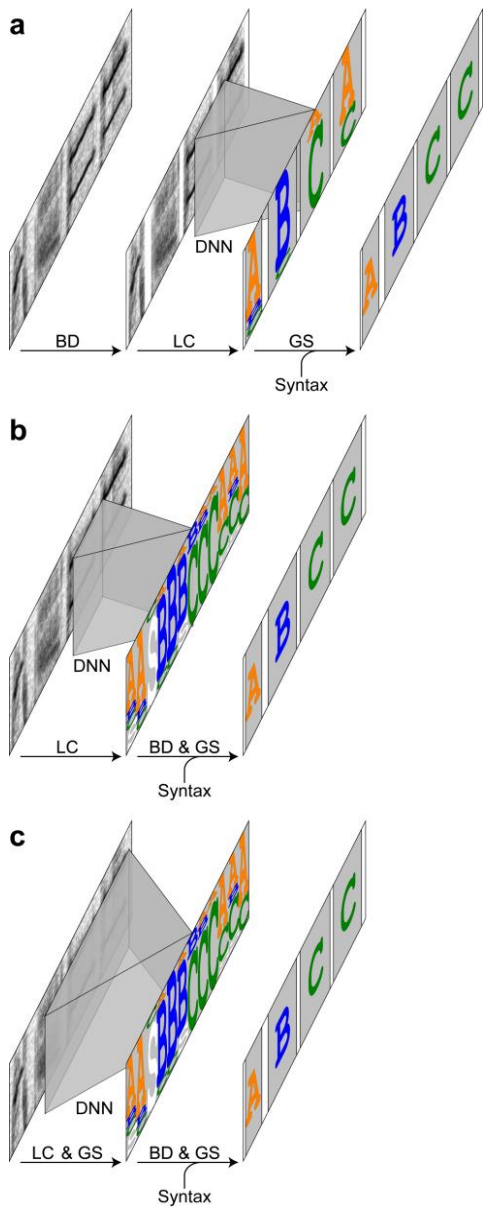

**Figure 1 | Three arrangements of methods for birdsong recognition.**

In this study we compared three arrangements of methods for birdsong recognition. (a) In the BD -> LC -> GS arrangement, first, note boundaries were detected by thresholding (BD step). The white regions indicate the detected inter-note silent intervals. Second, the spectrogram was locally classified with a DNN (LC step). The DNN outputs within the intervals of detected notes were averaged. The size of the colored letter indicates the magnitude of the DNN outputs. Finally, the classification results were combined with the syntax information with an HMM to compute the output sequences (GS step). (b) In the LC -> BD & GS arrangement, first, spectrograms within a sliding time window were locally classified with a DNN (LC step). Then boundary detection and global sequencing were simultaneously conducted with an HMM (BD & GS steps). The colored letters A, B, and C indicate the note classes, and the white letter S indicates the silent intervals. (c) In the LC & GS -> BD & GS arrangement, the input width of the DNN was longer than that in the LC -> BD & GS arrangement, hoping to capture the global syntax information with the DNN. The latter step was the same as the LC -> BD & GS arrangement.

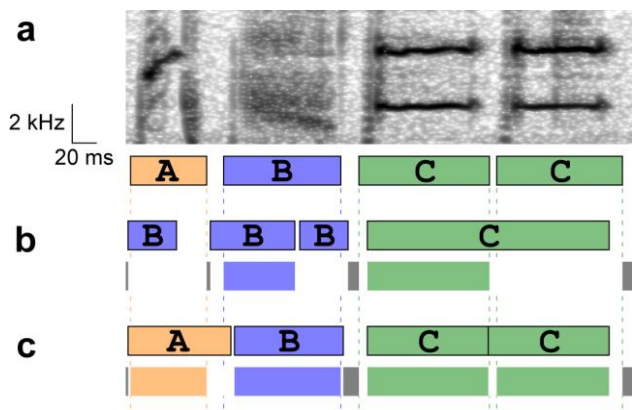

**Figure 2 | Evaluation with the matching error.**

To jointly evaluate the accuracy of note classification and boundary detection, we developed the new measure called the matching error. (a) An example spectrogram and the correct labels and boundaries. Note classes are indicated by letters. (b) An example recognition output (upper), and the correctly recognized intervals (lower). In the correctly recognized intervals, colored bars indicate the correctly recognized note intervals, longest overlaps with the correct label intervals. Gray bars indicate the correctly recognized silent intervals. The correctly recognized intervals appear to capture the performances properly even in such cases that a single note is recognized as two (notes B) and two notes are recognized as one (notes C). In both cases either of two overlapping intervals (the one with a longer overlap) was counted as matched intervals. There could be a case in which no matched interval is assigned (note A). The matching error is defined as the total length of the correctly recognized intervals divided by the length of the spectrogram, subtracted from one. (c) Another example showing the recognition outputs and the correctly recognized intervals with lower matching error.

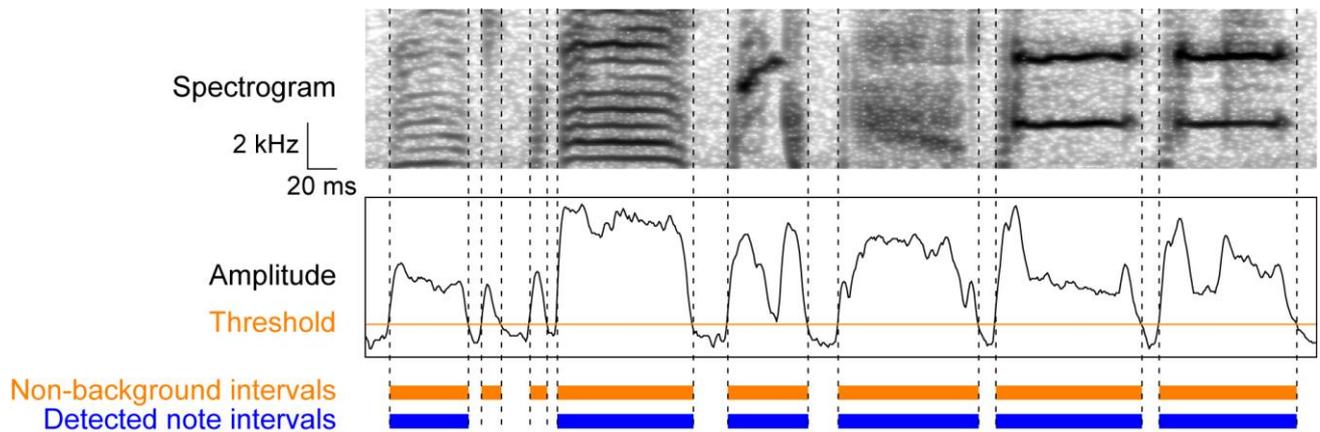

**Figure 3 | Boundary detection by thresholding.**

In the BD -> LC -> GS arrangement, note boundaries were detected using sound amplitude and interval duration. First, sound amplitude was computed as the sum of the logarithmic amplitude spectrum between the frequency band of 1 and 8 kHz in each 1 ms time bin of the spectrograms. Then, intervals in which amplitude was larger than a certain threshold were extracted as non-background intervals (orange bars). Finally, non-background intervals with duration shorter than a certain threshold were discarded. The remaining non-background intervals were considered as the putative note intervals (blue bars). These two thresholds were determined by cross-validation within training data sets.

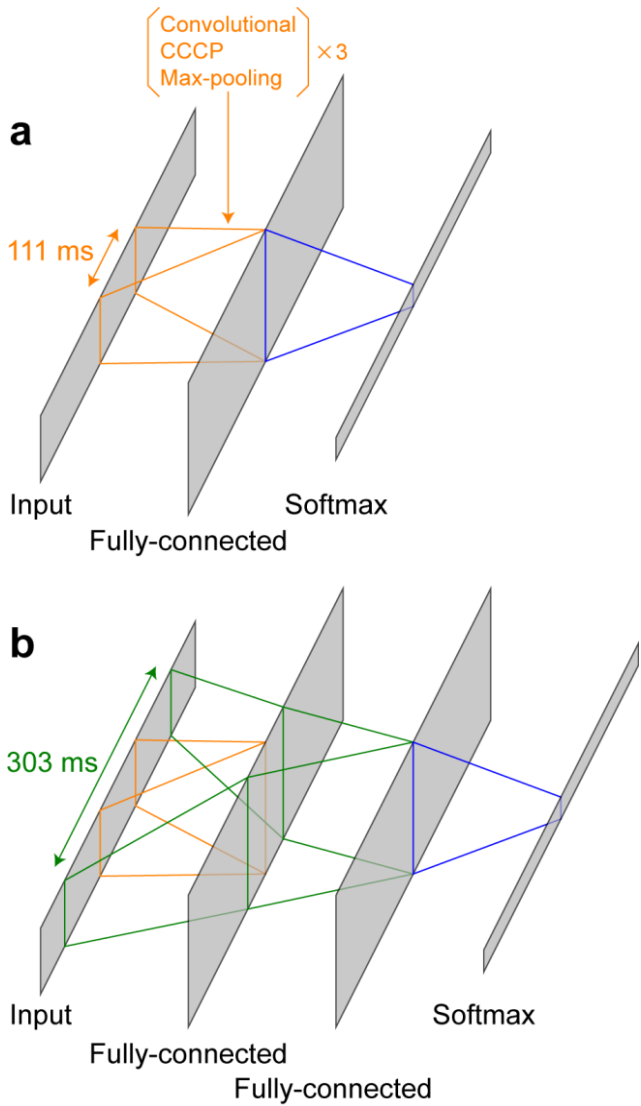

**Figure 4 | Local classification with a deep convolutional neural network.**

Local classification was conducted with a deep convolutional neural network. The spectrogram within a certain length of time window was classified into given classes with a DNN. (a) In the BD -> LC -> GS and the LC -> BD & GS arrangements, the length of the time window was 111 ms, roughly corresponding to the length of a single note. (b) In the LC & GS -> BD & GS arrangement, another fully-connected layer was inserted below the softmax layer of the network in (a), resulting in the length of the input time window being 303 ms, roughly corresponding to the length of three successive notes.

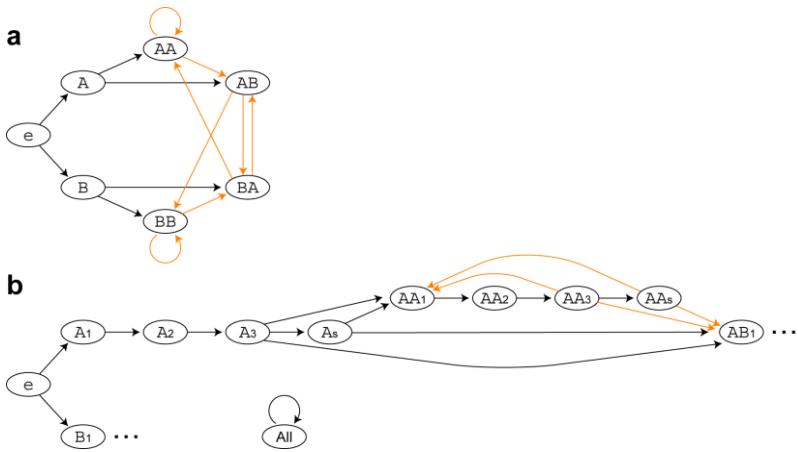

**Figure 5 | Syntax models used in HMMs.**

In the current study the song syntax in the Bengalese finches was modeled with a second-order Markov model. In this figure examples with two note classes (A & B) are shown. (a) A transition diagram in the BD -> LC -> GS arrangement. The initial state is indicated by the letter "e". The transition probabilities of orange arrows were computed from the training data sets. Those of black arrows were uniformly distributed (ie. all transition probabilities from states "e", "A", and "B" are 0.5). Sequence generation is allowed to stop at any states. (c) In the LC -> BD & GS and the LC & GS -> BD & GS arrangements, each state in (b) except the initial state was divided into four, first three of which corresponded to the note intervals and the last of which corresponded to the silent intervals. Transition to the next notes was allowed only from the third or fourth sub-states. All states had self-transitions. Since whole transition diagram would be too complex, only a part of it is shown in the figure.

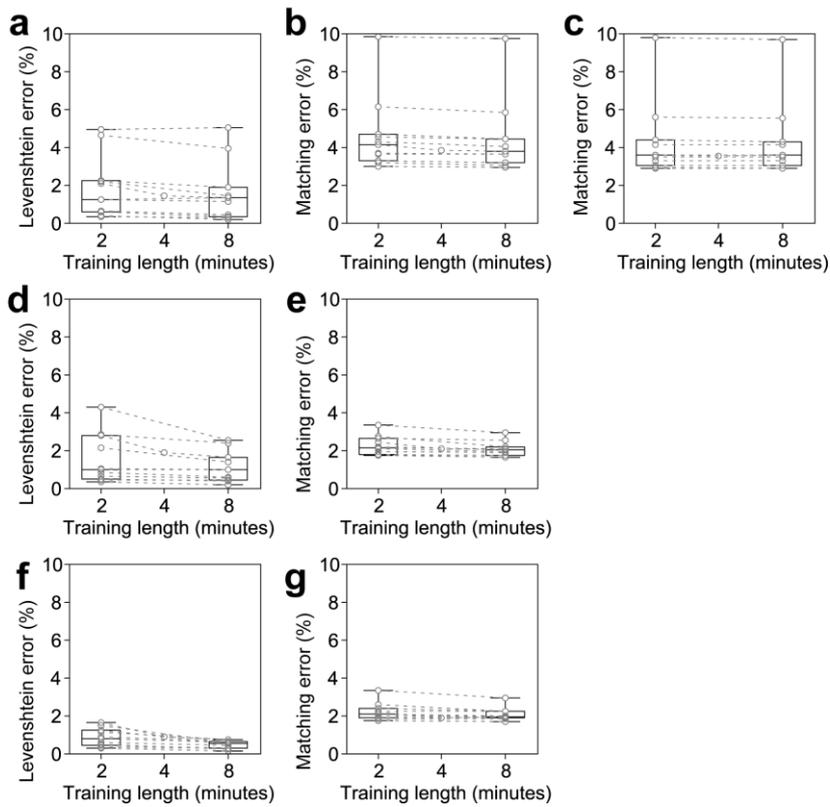

**Figure 6 | Validation errors.**

The Levenshtein errors (a, d, & f) and the matching errors (b, e, & g) of the results trained on 2, 4, and 8 minutes of training data sets in the BD -> LC -> GS arrangement (a & b), in the LC -> BD & GS arrangement (d & e), and in the LC & GS -> BD & GS arrangement (f & g). Medians, quartiles, minimums, and maximums were shown. Errors in each bird were indicated by open circles connected with dashed lines. (c) Matching errors ignoring note classes, evaluating the accuracy of boundary detection in the BD -> LC -> GS arrangement.

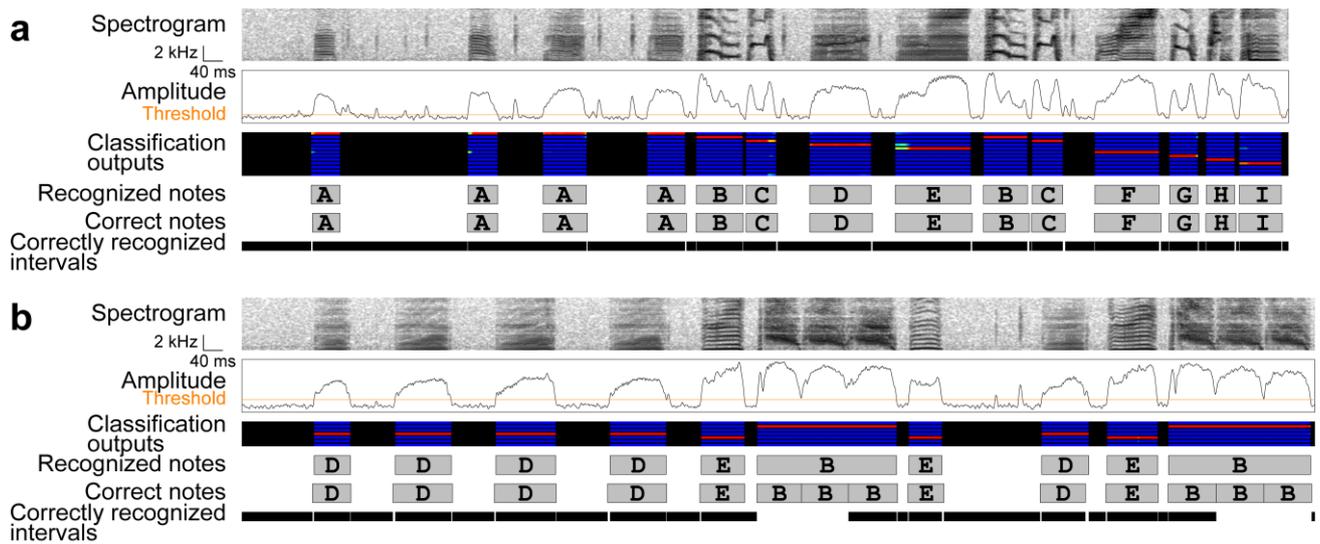

**Figure 7 | Recognition results in the BD -> LC -> GS arrangement.**

(a) A recognition result in one bird. From upper to lower: An input spectrogram, amplitude, outputs of local classification, recognized note intervals, correct note intervals, and correctly recognized intervals. Rows in the classification outputs correspond to the note classes. The black areas are putative silent intervals detected in the boundary detection step. Gray rectangles with letters indicate note intervals and classes. The correctly recognized intervals are indicated by black bars. (b) A result in another bird with poorer recognition accuracy.

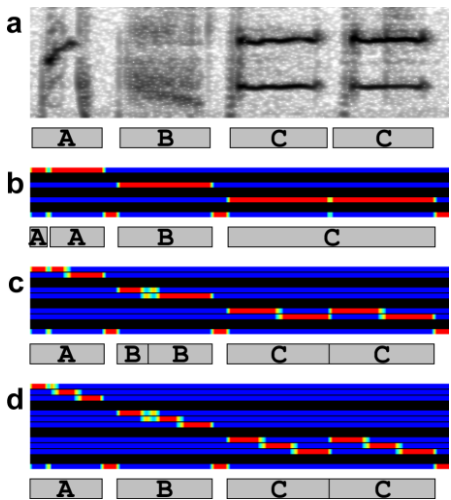

**Figure 8 | Sub-divisions in notes.**

In the LC -> BD & GS and the LC & GS -> BD & GS arrangements notes were divided into three parts for accurate boundary detection with an HMM. (a) An example input spectrogram (upper) and the correct note intervals (lower). Gray rectangles with letters indicate note intervals with note classes. (b) Example outputs of a DNN without sub-division in notes (upper) and the recognized sequence (lower). First three rows in the DNN outputs correspond to three note classes and the last to the class for background noise. If notes were not divided, small misclassifications of notes and background would lead to incorrectly divided notes (notes A) or connected notes (notes C). (c) Example outputs of a DNN with notes divided into two parts (upper) and the recognized sequence (lower). First six rows in the DNN outputs correspond to three note classes with two sub-divisions. The last row corresponds to the background noise. If notes were divided into two, small misclassifications of first and second sub-divisions would lead to incorrectly divided notes (notes B). The incorrect answers in (a) would be fixed in this configuration. (d) Example outputs of a DNN with notes divided into three parts (upper) and the recognized sequence (lower). First nine rows in the DNN outputs correspond to the three note classes with three sub-divisions. The incorrect answers in (b) would be fixed.

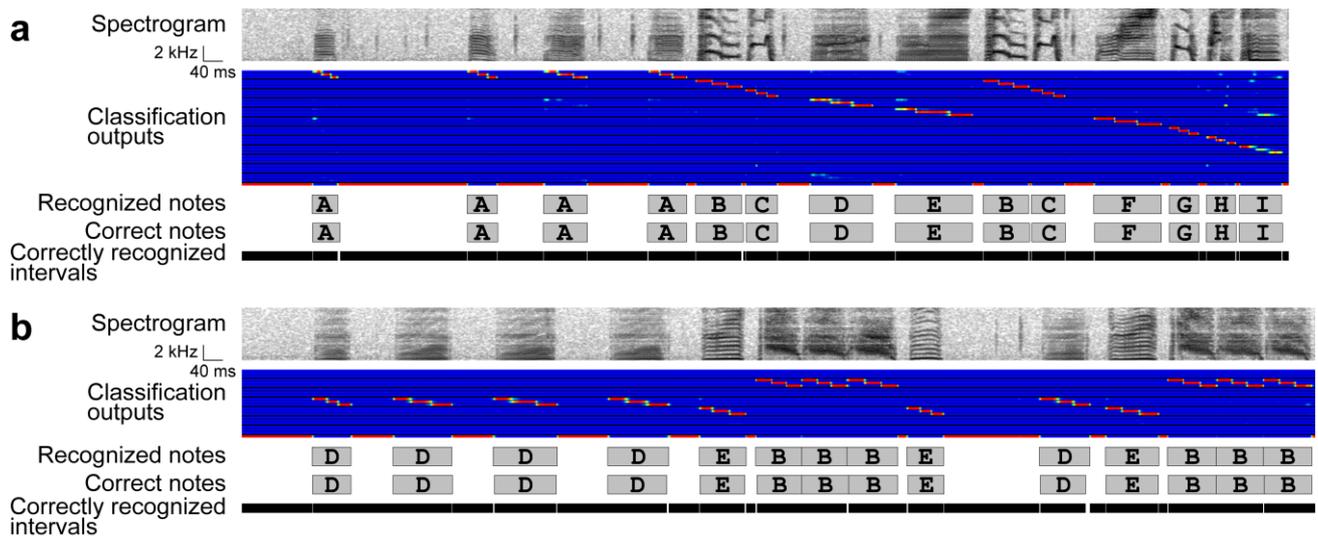

**Figure 9 | Recognition results in the LC -> BD & GS arrangement.**

(a) A recognition result in one bird. From upper to lower: An input spectrogram, outputs of local classification, recognized note intervals, correct note intervals, and correctly recognized intervals. Rows in the classification outputs correspond to twelve note classes with three sub-divisions. The bottom row indicates the class for the background noise. Gray rectangles with letters indicate note intervals and classes. The correctly recognized intervals are indicated by black bars. (b) A result in another bird.

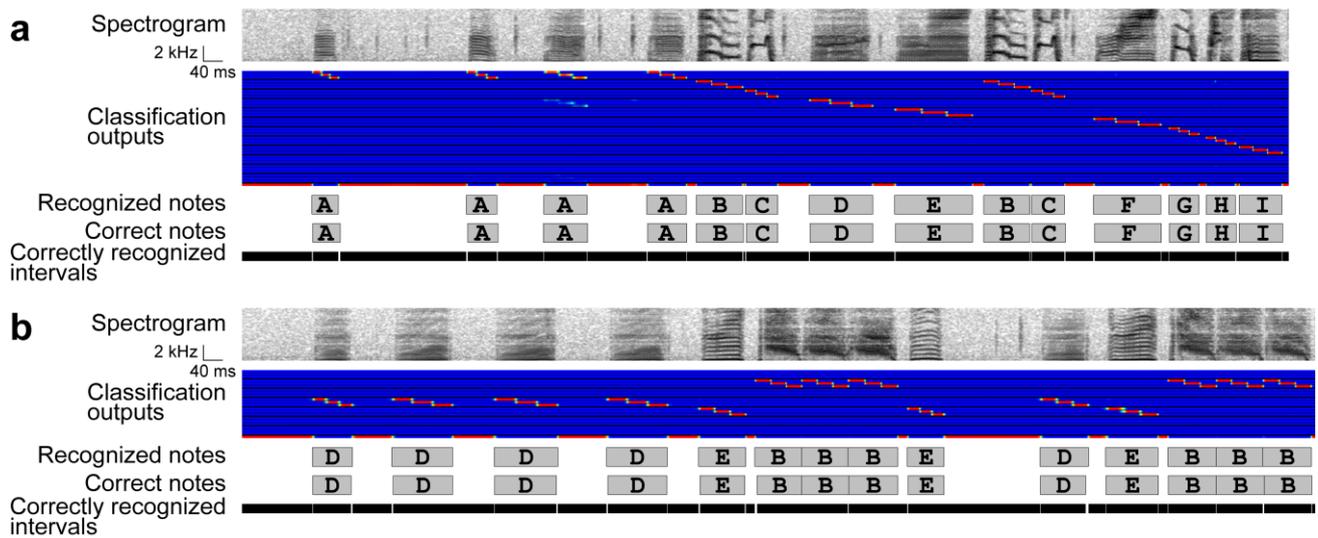

**Figure 10 | Recognition results in the LC & GS -> BD & GS arrangement.**

(a) A recognition result in one bird. From upper to lower: An input spectrogram, outputs of local classification, recognized note intervals, correct note intervals, and correctly recognized intervals. Rows in the classification outputs correspond to twelve note classes with three sub-divisions. The bottom row indicates the class for the background noise. Gray rectangles with letters indicate note intervals and classes. The correctly recognized intervals are indicated by black bars. (b) A result in another bird.

# Tables

|  | Levenshtein error | | Matching error | |
| :---: | :---: | :---: | :---: | :---: |
| Training length | 2 min | 8 min | 2 min | 8 min |
| BD -> LC -> GS | 1.58% | 1.33% | 4.41% | 4.28% |
| LC -> BD & GS | 1.45% | 1.06% | 2.14% | 1.97% |
| LC & GS -> BD & GS | 1.21% | 0.89% | 2.06% | 1.90% |

**Table 1 | Average Levenshtein errors and matching errors.**

**S1 Text. Data location and specification.**

All data used in this study are available at the following location:

https://zenodo.org/record/32024

The data contains multiple sound sequences produced by ten birds. Each sound sequence is stored in 16-bit linear PCM. The following table gives the specific description of the file format. First, the number of birds, the number of sound sequences, and the length of each sound sequence are sequentially stored. Then the contents of sound sequences follow. All values are stored in big-endian format.

| Byte offset | bytes | Description |
| --- | --- | --- |
| 0 | 4 | number of birds (N) |
| 4 | 4 | number of sound sequences in the first bird ($S_0$) |
| 8 | 4 | length of the first sequence in the first bird ($L_{0, 0}$) |
| 12 | 4 | length of the second sequence in the first bird ($L_{0, 1}$) |
| ……… | | |
| $8+4(N_0–1)$ | 4 | length of the last sequence in the first bird ($L_{0, S_0–1}$) |
| $8+4×N_0$ | 4 | number of sound sequences in the second bird ($S_1$) |
| $8+4N_0+4$ | 4 | length of the first sequence in the second bird ($L_{1, 0}$) |
| ……… | | |
| $4+4N+4\Sigma S_i–4$ | 4 | length of the last sequence in the last bird ($L_{N–1, S_{N–1}–1}$) |
| $4+4N+4\Sigma S_i$ | 2 | First value of the first sequence in the first bird |
| ……… | | |